\documentclass[10pt,cite,aps,pra,amsmath,amsfonts,amssymb,superscriptaddress,showpacs,twocolumn,nofootinbib]{revtex4}
\usepackage{graphics}
\usepackage{graphicx}
\usepackage{wasysym}
\usepackage{setspace}
\usepackage{color}
\usepackage{braket}

\renewcommand{\eqref}[1]{Eq.~(\ref{eq:#1})}

\newcommand{\mc}[1]{\mathcal{ #1 }}

\def\a{s}
\def\b{s}
\newcommand{\add}[1]{\if\a\b{{\color{red} #1}}\else{#1}\fi}
\newcommand{\comm}[1]{\if\a\b{{\color{blue}\{\small \sc #1\}}}\else{}\fi}
\newcommand{\del}[1]{{\if\a\b{{\color{magenta}[[#1]]}}\else{}\fi}}

\begin{document}

\title{Semi-Classical Field Theory as Decoherence Free Subspaces}

\author{Jaime Varela}
\affiliation{Berkeley Center for Theoretical Physics, Berkeley, CA 94720}
\affiliation{Lawrence Berkeley National Laboratory, Berkeley, CA 94720}

\begin{abstract}
We formulate semi-classical field theory as an approximate
decoherence-free-subspace of a finite-dimensional quantum-gravity
Hilbert space. A complementarity construction can be realized as a
unitary transformation which changes the decoherence-free-subspace.
This can be translated to signify that field theory on a global slice,
in certain space-times, is the simultaneous examination of two
different superselected sectors of a field theory.  We posit that a
correct course graining procedure of quantum gravity should be WKB
states propagating in a curved background in which particles exiting a
horizon have imaginary components to their phases.  The field theory
appears non-unitary, but it is due to the existence of approximate
decoherence free sub-spaces. Furthermore, the importance of operator
spaces in the course-graining procedure is discussed. We also briefly
touch on Firewalls.

\end{abstract}

\maketitle

\section{Introduction}
\label{sec:introduction}

The covariant entropy bound\cite{Bousso:1999xy} states that the Desitter
Hilbert space $\mathcal{H}_{\mathcal{A}}$ of area $\mathcal{A}$ has a
finite dimension given by:
\begin{equation}
\dim \mathcal{H}_{\mathcal{A}} = e^{\mathcal{A}/4}.
\end{equation}

Here $\mc{A}$ is the area of the Desitter horizon.  This finite
dimensional Hilbert space is in stark contradiction with the Hilbert
(Fock) spaces of quantum field theory which are always infinite
dimensional. One then must tread cautiously when attempting to import
field theory intuition to a finite dimensional quantum gravity (QG)
theory.  In our universe the Hilbert space dimension scales as
$e^{5.1\times 10^{121}}$.  The number of states corresponding to
semi-classical configurations parametrically behave as $e^{c A^{n}}$
where $n<1$ and $c$ is some constant (most likely $n=3/4$, see
\cite{tHooft:1993gx}).

This paper will be focused on the structure of quantum-computers which
simulate a field theory in curved space-time.  Qubits which give rise
to a classical gravitational field are treated as living in an
environmental Hilbert space sector while the matter lives in a system
Hilbert space (here the environment is also included as part of the
quantum computation). In order for matter to exhibit unitary evolution,
with its own system Hamiltonian, one needs to consider matter states in
decoherence-free-subspaces\cite{Lindarrev,Lindar,Lidar:1998hs}. Appendix
I. reviews the properties of decoherence free
subspaces \footnote{Throughout the rest of this paper we shall use Dfs
  to mean decoherence free subspace and DFS to mean the
  plural. Furthermore all DFS are only approximate unless otherwise
  specified.}.

We wish to consider quantum-computation simulations of field theory in
curved space-time because quantum-computers are
``UV-Complete''. Therefore, we can begin to model quantum gravity
effects by 1.) modifying the system environment interaction and 2.)
running the semi-classical simulation to the point where the Dfs
approximation becomes corrupted.  Quantum gravity effects can then be
related to environmental errors induced after sufficiently long times.
This work will be focused on the relationship between a semi-classical
field theory (field theory in curved space) and a quantum-computer
exhibiting decoherence-free-subspaces. The modeling of quantum gravity
effects will be the subject of future work.

Most of the discussion here will rely on the theory of decoherence,
quantum computing, and field theory in curved space-time.  In order to
make this paper more accessible, we present an intuitive description
of the proposal in the conclusions.



\section{Construction}

We consider a large but finite dimensional Hilbert space.  We claim
that the Hilbert space within a causal patch is given as:
\begin{equation}
\mathcal{H}= \mathcal{H}_{E} \otimes \mathcal{H}_{sc} =\mathcal{H}_E
\otimes \left( \mathcal{H}_s^{\perp} \oplus \mathcal{H}_{s} \right),
\end{equation}
where $\mathcal{H}_{sc}$ is the system, and $\mathcal{H}_{s}$ is a
decoherence free subspace of the quantum-gravity Hamiltonian $H$.
Furthermore, we postulate that there is a procedure, which we denote
by $\mathcal{S}$, which approximates the dynamics of $\mathcal{H}$
with a semi-classical field theory on a fixed background.  Namely:
\begin{equation}
\mathcal{S} \left(H,\mathcal{H}\right) \rightarrow \left(g_{\mu \nu}, \mathcal{H}_{\text{Fock}}\right)
\end{equation}
where $\mathcal{H}_{\text{Fock}}$ is the fock space associated with
particles propagating in a restricted space-time (For example, the
construction may not describe states propagating past a horizon.  It
is still not known why the space-time must be restricted.  Further
understanding of the gravitational sector is needed and is the subject
of further research.).

Also, there is a correspondence:
\begin{equation}
\mathcal{H}_{s} \rightarrow \mathcal{H}_{\text{Fock}}.
\end{equation}
It is unclear how this correspondence arises in the finite dimensional
Hilbert space $\mathcal{H}_{s}$.  This is related to the question of
how particles are approximated in a finite dimensional Hilbert space.
For this correspondence to work it is necessary for the operator
algebra of $\mc{H}_{sc}$ to allow Taylor expansion approximations in
terms of creation and annihilation operators\footnote{Here we will
  denote the operator space (equivalently operator algebra) of a
  Hilbert space $\mc{H}$ to be represented by the symbol
  $\mc{O}(\mc{H})$. Physically, this space represents the set of
  operators we will be working with as well as their commutation
  relations. We are including the commutator of any operator with the
  Hamiltonian when we speak of this space.}.  For example, the
Holstein Primakov transformation\cite{Holstein:1940zp} states that the
spin $S$ representation of the angular momentum operators is given by:
\begin{align}
S_{+} &= \sqrt{2 S} \sqrt{1-\frac{a^{\dagger} a}{2 S}} a ; \nonumber \\
S_{-} &= \sqrt{2 S} a^{\dagger} \sqrt{1-\frac{a^{\dagger}a}{2 S}} ; \nonumber \\
 S_{z} &= (S-a^{\dag} a).
\end{align}

For large $S$ the angular momentum operators are approximately bosonic
creation/annihilation operators.  How fermions may emerge is a more
complicated story and string net condensate models\cite{Levin:2004mi}
may provide an answer.

\section{Complementarity}

Complementarity\cite{Susskind:1993if} was initially proposed to
resolve a conflict between unitary black hole evaporation and
semi-classical field theory. Several analyses on a global space-like
slice, which extended through the event horizon, lead to the
conclusion that black holes clone quantum states.  Due to the fact
that the cloning of arbitrary states is a violation of quantum
mechanics, a complementary picture of black hole physics was
formulated.  In brief, complementarity states that black hole physics
can be described in the exterior or the interior but not both at the
same time.  A \textit{``complementarity transformation''} is a
transformation of the degrees of freedom (operators and states).
Initially the degrees of freedom describe physics in the exterior of a
black hole.  Once the transformation is applied, the degrees of
freedom describe physics in the interior.

Here we propose that the complementarity transformation should change
the properties of the Dfs. This can be realized as a unitary
transformation of the Hamiltonian which necessarily changes the
decoherence-free-subspace.  Applying the approximation $\mc{S}$ on the
new decoherence-free-subspace results in a new space-time $g_{\mu
  \nu}'$ and fock space $\mathcal{H}_{\text{Fock}}'$.  \textit{Field
  theory in curved space-time, on a global slice, can then be thought
  of the union of the two fock spaces and the gluing of the restricted
  space-times.} The motivation to consider changes in DFS as part of
the complementarity transformation is described in the next section.

  Before closing this section we comment on the relation of
  decoherence-free subspaces and super-selection sectors.  For certain
  Hamiltonians, see Appendix 2, states in DFS are grouped according to
  specific eigenvalues they posses (charge, for example).  Any
  superposition of states within the Dfs will evolve unitarily under
  the system Hamiltonian. Expressed in a field theory language, this
  is a statement of how superselection sectors decohere.  In other
  words, \textit{super-selection sectors arise through a systems
    interaction with the environment.}  In the case studied here the
  environment is the microscopic degrees of freedom which give rise to
  a gravitational field.

Field theory in curved space-time, on a global slice, can thus be seen
as the simultaneous examination of many super-selected sectors.  Thus,
one is improperly doing field theory in the sense described in
ref\cite{Streater:1989vi}.  The author however, prefers to view the
emergence of superselection sectors as a consequence of the Dfs
formulation.

\section{Physical Description of States}

There are many special cases which can be considered but here we focus
on the case where the Hamiltonian has only an approximate Dfs.  This
implies that there are no time-constants which vanish in the Dfs (see
Appendix 3).  If all time constants are positive, then unitarity will
appear to be violated for large enough time.

The mapping $\mc{S}$ then takes the states in these classes of DFS to
WKB modes propagating in a curved background.  Furthermore, all modes
will have a positive imaginary component to their phases\footnote{The
  convention we use is such that a positive imaginary part of the
  phase will result in a decaying exponential.}.  Such non-unitary
field theories can be constructed on a curved background.  However,
such theories have been ruled out in the past due to a violation of
unitarity.  \textit{This apparent violation of unitarity is justified
  in the Dfs framework}.

We conjecture that Hamiltonians which exhibit only approximate DFS are
mapped to those which have initial data on a particular Cauchy
surface.  This Cauchy surface is one which the congruence of future
directed geodesics emanating from it become inextensible in finite
affine parameter (a singularity). Hamiltonians with exact Dfs are in
another extreme.  We conjecture that these Hamiltonians describe
space-times with no singularities\footnote{The exterior picture in
  black hole evaporation has an interesting WKB structure. Here the
  imaginary component to the phases has a negative sign (this
  corresponds to particles being created).  Furthermore, in the
  exterior picture there are a large set of subsystems whose time
  constants go to zero.}

 It may seem strange that we are discussing WKB theories with
 non-unitary evolution when we expect our universe to evolve unitarily
 (non-unitarity in the WKB theory is seen in the non-conservation of a
 probability amplitude).  The reader should recall however, that we
 are using WKB only as an effective theory in which we ignore the
 dynamics of the gravitational environment.  The reduced density
 matrix of the matter sector need not evolve unitarily after the trace
 is preformed\cite{Schlosshauer(2007)}.  If we do not perform the
 trace, and keep track of the gravitational sector, then the full
 theory is unitary. An excellent discussion on the evolution of the
 reduced density matrix for black hole evaporation can be found in
 ref\cite{Anglin:1994yi}.




\section{Comment on the Procedure $\mathcal{S}$}

So far, this discussion has relied on the mysterious ``procedure''
called $\mc{S}$.  Here we do not construct the procedure but we
discuss how it must be viewed in quantum information theory terms.

A quantum computation can be viewed as a selection of the quintet:
\begin{equation}
\left(\tau, H,\mathcal{H},\mathcal{O}(\mathcal{H}), | \psi \rangle \right),
\end{equation}
where $H$ is the Hamiltonian, $\mathcal{H}$ is the Hilbert space,
$\mathcal{O}(\mathcal{H})$ is the relevant operator algebra, $| \psi
\rangle$ is a state of the Hilbert space, and $\tau$ is time.  The
correspondence of these objects to quantum information theory is as
follows:
\begin{align}
\tau &\rightarrow \text{Parameter of evolution gate} \nonumber \\
H &\rightarrow \text{A generator for the evolution gate} \nonumber \\
\mc{H} &\rightarrow \text{Qubit Hilbert Space} \nonumber \\
\mc{O}(\mc{H}) &\rightarrow \text{Set of gates and/or generators of gates} \nonumber \\
| \psi \rangle &\rightarrow \text{Initial Qubit Configuration}.
\end{align}
By evolution gate we simply mean $e^{i H \tau}$, the standard time
evolution operator\footnote{Objections can be made that the true
  quantum gravity state should be
  static\cite{DeWitt:1967yk,Nomura:2012zb}.  However
  one can always discuss ``Evolution Without Evolution'' by coupling to
  an external system\cite{Page:1983uc}.  Thus, we ignore the possibility of a
  static state.}.

Focusing on the Dfs, we can see that it too has such a quintet.  Namely:
\begin{equation}
\left(\tau, H_{sc},\mathcal{H}_{sc},\mathcal{O}(\mathcal{H}_{sc}), | \psi_{sc} \rangle \right),
\end{equation}
where $H_{sc}$ is the Hamiltonian of the system, and with similar
definitions for the other elements of the quintet\footnote{It is
  essential that we are focusing on the Dfs.  If we did not, then the
  evolution gate would be error prone.}. The quantum computation
defined by this quintet should realize a simulation of a field theory
in curved space\footnote{By simulation, we mean a circuit construction
  which takes the initial qubit configuration to a particular
  out-state.  The out-state can be used to obtain expectation values
  of interest (N-point functions). See ref\cite{Jordan:2011ne} for a
  simulation of $\phi^4$ theory.}.

Aside from the quantum simulation requirement there will also be a
formal tracing (or averaging) procedure over the environmental Hilbert
space $\mc{H}_{E}$.  Diffeomorphism invariance can be understood as
the ability to select the operator algebra $\mc{O}(\mc{H}_{E})$
arbitrarily (or within some reasonable constraint).

Let the set $(\gamma_{ij}(I),\pi_{ij}(J))$ be the generators of
$\mc{O}(\mc{H}_{E})$, where $(i,j,I,J)$ are indices\footnote{Readers
  familiar with the ADM formalism\cite{ADM,DeWitt:1967yk} will
  recognize the notation}. These generators must have approximate
bosonic algebras.  Furthermore, the commutators of these operators
with the Hamiltonian must have an Erhenfest approximation which
reproduces a discretized version of the first order differential
equations of General Relativity\footnote{This is a purely mathematical
  question.The existence (non-existence) of these algebraic structures
  will validate (invalidate) this work.}.

In other words, \textit{the selection of the operator algebra defines
  the lapse and shift functions in the ADM formulation of General
  Relativity}. The study of non-linear algebras is poorly understood
(at least in physics) and the Author does not know if this methodology
can be realized in finite dimensional Hilbert spaces.  If it can not,
it would be interesting to examine which mathematical spaces do allow
such algebraic structures.

Furthermore, the method described above may not be the most practical in
obtaining Einsteins equations from a many-body quantum system.  Tensor
network methods may be more practical to discuss the emergence
of space-time as they have already been shown to be of use in this endeavour\cite{Swingle:2009bg,Swingle:2012wq}.

\section{Discussion on Firewalls}

Here the Author does not claim to solve the firewall paradox.  This
paradox was first posed in ref\cite{SamB}, and later expanded upon in
ref\cite{Almheiri:2012rt}.  We shall attempt to formulate this paradox
in an information theoretic way, keeping decoherence-free subspaces in
mind.

The firewall problem can be recast into the statement: \textit{``Can a
  semi-classical observer extract a purification of a bit from the
  early Hawking radiation''}\footnote{The Author would like to thank
  Raphael Bousso for making this clear to him.}. By ``semi-classical
observer'', we mean an observer which only has control over gates
which act on his/her Dfs. The number of independent ``semi-classical
gates'' scale as $\left( \dim \mc{H}_{sc} \right)^2$, which is far
smaller than the total number of gates in the system.  We restrict the
number of gates a semi-classical observer can carry because gates
themselves are physical objects.  Therefore, packing a large number of
gates into a small region will result in significant back-reaction to
the geometry. One can also imagine doing a computation very far away
and then jumping into the black hole.  Wether or not someone can see a
firewall is a question on the time-scale restrictions on quantum
computations. We will not discuss such thought experiments here.

Given the restriction on the number of gates, the firewall question
becomes:

\begin{center}
\textit{In the timescale of interest, can an infalling observer
  perform a quantum computation of Firewall Complexity using only
  semi-classical gates.}
\end{center}

By ``Firewall Complexity'' we mean the quantum complexity class in
which the purification computation belongs\footnote{Several authors
  have discussed attaching physical significance to complexity
  classes\cite{Harlow:2013tf,Susskind:2014rva}.  This phenomenon
  appears here as well.}. The firewall question is a purely quantum
information theory question and should have an answer, at least once
we know exactly what computation we need to do.

The author conjectures that there is no firewall.  The reason being that
one expects a black hole to be a quantum object
which preforms operations, on infalling matter, which simulate
general relativity.  This is simply a conjecture and the question
still needs to be answered.  If firewalls do exist however, they could be described as regimes in the quantum computation in which a subsystem of interest no longer evolves unitarily under its own system Hamiltonian (how this is done and the physical interpretation of such a procedure is a matter of further work).

\section{Conclusions}

Here we take a brief aside to discuss the philosophical implications
of this formalism.  If this formalism describes nature then it implies
that matter states are living in a decoherence-free-subspace in a full
quantum gravitational theory.  Viewed computationally, matter are
states living in an approximately error-free register of a quantum
computer.

Furthermore, this framework makes a statement on the fundamental
symmetry group of nature.  If we are a finite dimensional Hilbert
space then the Poncaire group can't be the fundamental group since
there are no finite unitary representations of the Lorentz group.
Quantum computations in finite dimensional Hilbert spaces may describe
theories with emergent Lorentz symmetry, such as those described in
ref\cite{Horava:2009uw,Levin:2004mi}.  Further study of these theories
and their relationship with quantum computation would be interesting.

Finally, it should be mentioned that the role decoherence has to play
in quantum gravity has been emphasized by many
authors\cite{Zurek:1982zz,Anastopoulos:2008xz,Anastopoulos:2007xv,Nomura:2012nt,Nomura:2012cx,Gambini:2012tha}.
With pioneering work done by Kiefer and Joos\cite{Kiefer:1997hv}, and
Anglin, Laflamme, Zurek, and Paz\cite{Anglin:1994yi}.  This work hopes
to add to the above body of work in a small way.

\section{Acknowledgments}

The Author would like to thank Zachary Fisher, Raphael Bousso, Philipp
Dumitrescu, Nathan Haouzi, Daniel Varjas, Sean Weinberg and Yasunori
Nomura for excellent discussions. I also acknowledge NSF grant number
DGE-1106400 and UC Berkeley's Chancellor Fellowship.

\section*{Appendix 1: Decoherence Free Subspaces and Subsystems}
\label{sec:appendix}

In this appendix we review some aspects of decoherence free subspaces
and subsystems.  For a more thorough review the reader should
consult\cite{Lindarrev,Lindar}. The Hilbert space is given by:

\begin{equation}
\mathcal{H}= \mathcal{H}_E \otimes \mathcal{H}_{sc}
\end{equation}

The Hamiltonian is:
\begin{equation}
H = H_E+H_{sc} + H_{int}
\end{equation}
where $H_E$ acts only on $\mathcal{H}_E$, $H_{sc}$ acts only on
$\mathcal{H}_{sc}$, and $H_{int}$ is an interaction Hamiltonian.  A
subspace $\mathcal{H}_s \subset \mathcal{H}_{sc}$ is a
decoherence-free-subspace, with precision $\epsilon$, if any density
matrix $\rho_s$ with support in $\mathcal{H}_{s}$ satisfies:
\begin{equation}
\rho_s(t) = e^{i H_{sc} t} \left( \rho_{s} \right) e^{-i H_{sc} t} + \mathcal{O}(\epsilon).
\end{equation}
where:
\begin{equation}
\rho_s(t) = \text{tr}_E \left( e^{i H t} \rho_E \otimes \rho_{s} e^{-i H t} \right)
\end{equation}

for a large set of $\rho_E$.  By $\mathcal{O}(\epsilon)$ we mean that
any expectation value of an operator has only order $\epsilon$
corrections.

It is worth noting that the identification of the
decoherence-free-subspace is highly dependent on the Hamiltonian.
Furthermore, the existence of exact decoherence-free-subspaces does
\textbf{not} necessarily imply that $H_{int}=0$.

The above definition is equivalent to the standard way of defining
decoherence-free-subspaces but it is usually not the most practical.
The above definition is useful however, because it allows one to work
backwards and construct decoherence-free-subspaces.  Thus, it is
independent on whether $\mathcal{H}_{sc}$ has many
subsystems. \textit{Constructing decoherence-free-subspaces, in
  situations in which the relevant Hilbert space has many subsystems,
  is done by partitioning states based on the evolution of the subsystems.}

\section*{Appendix 2: Operator Spaces and DFS}
\label{sec:appendix2}

There is a relationship between Lie Groups and DFS in special cases
where the Hamiltonian possesses a certain structure.  The discussion
here follows closely Theorem 1 of ref\cite{Lidar:1998hs}.  Here we
state the theorem without proof.

Let the interaction Hamiltonian be given by:
\begin{equation}
H_{int} = \sum_{\alpha} \mathcal{B}_{\alpha} \otimes \mathcal{F}_{\alpha},
\end{equation}
where $\mc{B}_{\alpha}$ act only on $\mathcal{H}_{E}$ and
$\mc{F}_{\alpha}$, called error generators, act on $\mathcal{H}_{sc}$.
The decoherence-free-subspaces are those which posses degenerate
eigenvalues with respect to all the error generators.  In other words,
the set $|i\rangle$ which satisfy $\mathcal{F}_{\alpha} |i\rangle =
c_{\alpha} |i \rangle$, for all $\alpha$, form a Dfs.

If the $\mc{F}_{\alpha}$ form an $M$ dimensional semi-simple lie
algebra in the $N$ dimensional matrix representation, where $N =
\text{dim} \mathcal{H}_{sc}$, then the \textit{decoherence free
  subspaces live in one dimensional irreducible representations of the
  lie-group} (singlet states).

\section*{Appendix 3: DFS and Hilbert Spaces with many subsystems}
\label{sec:appendix}

Here we discuss how to view a Dfs when $\mc{H}_{sc}$ contains many
subsystems.

If the Hilbert space is given as:
\begin{equation}
\mathcal{H} = \mathcal{H}_E \otimes \left[ \large\otimes_i
  \mathcal{H}_i \right]
\end{equation}

Then we can define:
\begin{equation}
\rho_j (t) = \text{tr}_{\mathcal{H}_E} \left[\text{tr}_{i\neq j}
  \left( e^{i H t} \rho e^{-i H t} \right) \right].
\end{equation}

Here $H$ is the total Hamiltonian. In many cases the form of the
density matrix is:
\begin{equation}
\label{eq:decay}
\rho_j(t) \propto e^{-\alpha_j t}
\end{equation}
with some time constant $\alpha_j$.  Here we are using a shorthand
notation to denote the strength of the non-unitary behaviour of the
density matrix evolution. More formally, one can think about the
evolution of the expectation values of all physically relevant
operators.  If all expectation values have decay behaviour equal or
worse than that described in eq~\ref{eq:decay}, then the time constant
can be defined. We will be using the time constant as a measure of the
breakdown of unitarity. (We remind the reader that density matrix
evolution need not be unitary when a environmental trace is preformed.
Also, it should be noted that the above measure of non-unitarity holds
only in a free theory where $i$ and $j$ do not interact heavily. A
better definition in which the subsystems interact is needed.)

Decoherence free subspaces may be defined as the density matrices in
which the maximum time constant of the set of all time constants is
below some threshold.  In other words $\text{Max}\left( | \alpha_i |
\right) < \alpha_{cut}$.

For exact DFS spaces we expect $\text{Max}( | \alpha_i | ) \rightarrow
0$. What I described above is only an approximate way of defining Dfs
in the product Hilbert space. Also, this requires understanding on the
initial factorization of the environment.  However, when discussing
decoherence, some knowledge of the environment is always required.






\end{document}